\documentstyle[12pt,epsfig]{article}
\def\ga{\overline{g}_{\mbox{\tiny A}}}
\def\gv{\overline{g}_{\mbox{\tiny V}}}
\def\gp{\overline{g}_{\mbox{\tiny P}}}
\def\gpa{\overline{g}_{\mbox{\tiny P1}}}
\def\gpb{\overline{g}_{\mbox{\tiny P2}}}
\def\gw{\overline{g}_{\mbox{\tiny W}}}

\def\gA{g_{\mbox{\tiny A}}}

\def\gV{g_{\mbox{\tiny V}}}
\def\gM{g_{\mbox{\tiny M}}}

\def\gP{g_{\mbox{\tiny P}}}

\def\lsim{\:\raisebox{-0.5ex}{$\stackrel{\textstyle<}{\sim}$}\:}
\def\gsim{\:\raisebox{-0.5ex}{$\stackrel{\textstyle>}{\sim}$}\:}

\def\be{\begin{equation}}
\def\ee{\end{equation}}
\def\br{\begin{eqnarray}}
\def\er{\end{eqnarray}}
\def\brn{\begin{eqnarray*}}
\def\ern{\end{eqnarray*}}
\def\E {{{\cal E}}}
\def\T {{{\cal T}}}
\def\ie{{\em i.e., }}
\def\nn{\nonumber }
\def\rf#1{{(\ref{#1})}}
\def\sss{\scriptscriptstyle}

\def\kb {{\bf k}}
\def\pb {{\bf p}}
\def\qb {{\bf q}}

\def\rb {{\bf r}}
\def\sq{\sqrt{2}}
\def\d{\dagger}
\def\g {{\gamma}}
\def\s {{\sigma}}

\def\l {{\lambda}}
\def\go{\rightarrow}
\def\mbs{\mbox{\boldmath$\sigma$}}
\def\Jb{ {\bf J}}
\def\x{\times}
\def\ket#1{|#1 \rangle}

\def\Ket#1{||#1 \rangle}
\def\Bra#1{\langle #1||}
\def\ov#1#2{\langle #1 | #2  \rangle }
\def\M {{{\cal M}}}
\def\rh {\hat{r}}
\def\kh {\hat{k}}
\def\zh {\hat{z}}
\def\threej#1#2#3#4#5#6{\left(\negthinspace\begin{array}{ccc}
#1&#2&#3\\#4&#5&#6\end{array}\right)}
\def\del {\delta}

\def\etal{{\it et al. }}

\begin{document}

\title{RPA Puzzle in ${^{12}C}$ Weak Decay Processes}

\author{ F. Krmpoti\'{c}, A. Mariano and A. Samana \\
{\it Departamento de F\'{\i}sica, Facultad de Ciencias Exactas,} \\
{\it Universidad Nacional de La Plata,} \\ {\it C.C. 67,
1900 La Plata,
Argentina}}
\maketitle

\begin{abstract}
We explain the origin of the difficulties that appear in a straightforward application
of the QRPA in ${^{12}C}$, and we demonstrate that it is imperative to use the projected
QRPA (PQRPA).
Satisfactory results, not only for the weak processes among  the ground states of the triad
$\{{{^{12}B},{^{12}C},{^{12}N}}\}$, but also for the inclusive ones  are obtained.
We sketch as well  a new formalism for the neutrino-nucleus interaction that furnishes
very simple final formulae for the muon capture rate and neutrino induced cross sections.
\end{abstract}

Pacs: 13.75.Gx, 14.20.Gk, 13.40.Em

Keywords: neutrino-nucleus cross section, muon capture, beta decay,projected
QRPA \vskip 1.cm


New types of nuclear weak
processes have been measured  in recent years. They are based on neutrino
and antineutrino  interactions
with complex nuclei and, rather than being used to study the
corresponding cross sections,  they are mainly aimed to inquire on
possible exotic properties of neutrino themselves, such as
neutrino oscillations and the associate neutrino massiveness,
which  are not contained in the Standard Model (SM) of elementary
particles.

So, in  neutrino oscillation experiments with   liquid scintillators,
the charge-exchange reactions ${^{12}C}(\nu_e ,e^-){^{12}N}$ and
${^{12}C}(\nu_\mu ,\mu^-){^{12}N}$,  both {\em exclusive} (to the $1^+$ ground state) and
{\em inclusive} (to all final states), are just  tools. As such, and  to be useful,
the corresponding cross sections $\sigma_{e,\mu}^{exc}$ and $\sigma_{e,\mu}^{inc}$ must be
accurately accounted for by nuclear structure models.

>From the recent works \cite{Kol94,Aue97,Kol99,Vol00} we have learned, however, that
neither  RPA nor  QRPA are able to explain  the weak processes
($\beta$-decays, $\mu$-capture, and neutrino  induced reactions)
among the ground states of the triad $\{{{^{12}B},{^{12}C},{^{12}N}}\}$.
In fact,  in the RPA a rescaling factor of the order of $4$ is needed
to bring the calculations and data in agreement \cite{Kol94}, and  a
subsequent {\it ad hoc} inclusion of partial occupancy of the
$p_{1/2}$ subshell reduces this factor to less than $2$ \cite{Aue97,Kol99}.
But, when the RPA is supplemented with the pairing correlations in a self-consistent
way, \ie  in the framework of a full QRPA \cite{Vol00}, the
exclusive cross sections turn out to be again overestimated  by a factor of $\cong 4$.
 Moreover, Volpe {\it et al} \cite{Vol00} have called attention
to ``{\it difficulties in choosing the ground state of $^{12}N$ because
the lowest state is not the most collective one}" when the QRPA is used.


In the present  paper we explain the origin of the difficulties that appear
in a straightforward application of the BCS approximation in a light nucleus
such as ${^{12}C}$, and we demonstrate that the problem is circumvented by
the employment of the particle number projected BCS (PBCS).
We show simultaneously that the proton-neutron QRPA is not a
recommended approach, and that the aforementioned RPA puzzle is solved
within the projected QRPA (PQRPA) for the charge-exchange excitations \cite{Krm93}.
The later approach furnishes satisfactory results not only for the weak processes
among  the ground states of the triad $\{{{^{12}B},{^{12}C},{^{12}N}}\}$, but also
for the inclusive weak processes.
For numerical evaluation of the weak decay observables we have found it suitable
to develop a new theoretical framework, which is similar to that
build  up by Barbero {\it et al.} \cite{Bar98} for the neutrinoless double beta
decay. The motivation for that and the complete formulation will be exposed elsewhere.
Here we just explain the notation and exhibit the final formulae.

The weak Hamiltonian is expressed in the form \cite{Bar98,Tom91,Doi93}
\br
H_{{\sss {W}}}(\rb)&=&\frac{G}{\sq}J_\mu^\d L^\mu(\rb)+h.c.,
\label{1}\er
where
\br
J_\mu&=&\g_0
\left[\gV\g_\mu+\frac{\gM}{2M}i\s_{\mu\nu}k_\nu
-\gA\g_\mu\g_5+\frac{\gP}{m_\ell} k_\mu\g_5\right],
\label{2}\er
is the hadronic current operator, and
\br
L_\mu(\rb)&=&\overline{u}_{s_\ell}(\pb,E_\ell)\g_\mu(1-\g_5)u_{s_\nu}(\qb,E_\nu)e^{i\rb\cdot
\kb}
\label{3}\er
is the plane waves approximation for the matrix element of the leptonic current;
$G=(3.04545\pm 0.00006){\times} 10^{-12}$ is the Fermi coupling constant (in natural units)
\cite{Tow95},
\br
k = P_i-P_f\equiv \{k_0,\kb\}
\label{4}\er
is the momentum transfer ($P_i$ and $P_f$ are momenta of the initial and final nucleon
(nucleus)), $M$ is the nucleon mass, $m_\ell$ is the mass of the charged lepton,
and $g_{\sss V}$, $g_{\sss A}$, $g_{\sss M}$ and $g_{\sss P}$ are, respectively, the vector,
axial-vector, weak-magnetism and pseudoscalar effective dimensionless coupling constants.
Their numerical values are \cite{Tom91,Doi93,Tow95}:
\br
g_{\sss V}&=&1;~
g_{\sss A}=1.26;~
g_{\sss M}=\kappa_p-\kappa_n=3.70;~
g_{\sss P}=g_{\sss A}\frac{2M m_\ell }{k^{2}+m_\pi^2}.
\label{5}
\end{eqnarray}
The above estimates for $g_{\sss M}$ and $g_{\sss P}$ come from the (well tested) conserved
vector current (CVC) hypothesis, and from the partially conserved axial vector current
(PCAC) hypothesis,  respectively. In the numerical calculation we will use an effective
axial-vector coupling
$g_{\sss A}^{\sss}=1$ \cite{Bro85,Cas87,Ost92}.
The finite nuclear size (FNS) effect is incorporated via the dipole form factor
with a cutoff $\Lambda=850$ MeV, \ie as \cite{Kur90}:
\be
g\go g\left( \frac{\Lambda^{2}}{\Lambda^{2}+k^{2}}\right)^{2}.
\label{6}\ee

To use \rf{1} with  the non-relativistic nuclear wave
functions, the Foldy-Wouthuysen transformation has to be performed
on the hadronic current \rf{2} . When the velocity dependent
terms are neglected
\footnote{ The effect  of the nucleon-velocity terms is of the order of a few per cent,
in both the neutrino-nucleus scattering \cite{Kur90} and in the muon capture\cite{Luy63,Kuz01}.}
, this yields \cite{Bli66}:
\br
J_0&=&\gV-(\ga+\gpa)\mbs\cdot\hat{\kb},
\nn\\
\Jb&=&\gA\mbs-i\gw\mbs\x\hat{\kb}-\gv\hat{\kb}-\gpb(\mbs\cdot\hat{\kb})\hat{\kb},
\label{7}\er
where the following short notation has been introduced:
\br
\gv&=&\gV\frac{|\kb|}{2M};~
\ga=\gA\frac{|\kb|}{2M};~
\gw=(\gV+\gM)\frac{|\kb|}{2M},
\nn\\
\gpa&=&\gP\frac{|\kb|}{2M}\frac{k_0}{m_\ell};~
\gpb=\gP\frac{|\kb|}{2M}\frac{|\kb|}{m_\ell}.
\label{8}\er

For the neutrino-nucleus reaction $k=p-q$, with $p\equiv\{E_\ell,\pb\}$
and $q\equiv\{E_{\nu},\qb\}$, and the corresponding cross section from the initial
state $\ket{J_i}$ to the final state $\ket{J_f}$ reads 
\br
\s(E_\ell,J_f)& = &\frac{|\pb| E_\ell}{2\pi} F(Z+1,E_\ell)
\int_{-1}^1
d(\cos\theta)\T_{\s}(|\kb|,J_f),
\label{9}\er
where $F(Z+1,E_\ell)$ is  the Fermi function, $\theta\equiv \hat{\qb}\cdot\hat{\pb}$, and
\br
\T_{\s}(|\kb|,J_f)\equiv
\frac{1}{2J_i+1}\sum_{{s_\ell},M_i}\sum_{{s_\nu},M_f}
\left|\int d\rb\psi^*_f(\rb)H_{{\sss {W}}}(\rb)\psi_i(\rb)\right|^{2},
\label{10}\er
with  $\psi_i(\rb)\equiv \ov{\rb}{J_iM_i}$ and $\psi_f(\rb)\equiv \ov{\rb}{J_fM_f}$
being the nuclear wave functions.
The transition amplitude  can be cast in the form:
\br
\T_{\s}(|\kb|,J_f)&=&G^{2}\left(\M_{{\sss {V}}}K_{{\sss {V}}}+
\sum_{\mu=-1,0,+1}\M_{{\sss {A}}}^\mu K_{{\sss {A}}}^{\mu}\right),
\label{11}\er
where
\br
\M_{{\sss {V}}}&=&\frac{4\pi}{2J_i+1}\sum_{J}
\left|\Bra{J_f}i^{J}j_J(|\kb|r)Y_{J}(\rh)\Ket{J_i}\right|^2,
\nn\\
{\M}_{{\sss {A}}}^{\mu}&=&\frac{4\pi}{2J_i+1}\sum_{J}\left|
\sum_{L}\sqrt{2L+1}\threej{L}{1}{J}{0}{-\mu}{\mu}
\Bra{J_f}i^{L}j_L(|\kb|r)\left[Y_{L}(\rh)\otimes\mbs\right]_{J}\Ket{J_i}\right|^2,
\label{12}\er
are the nuclear matrix elements, and
\br
K_{{\sss {V}}}&=&g_{{\sss {V}}}^2L_{4}+
2g_{{\sss {V}}}\overline{g} _{{\sss {V}}}L_{40}
+\overline{g} _{{\sss {V}}}^2L_{0}
\nn\\
K_{{\sss {A}}}^{\mu}&=&\left\{
\begin{array}{ll}
  \left(g_{{\sss {A}}}-\overline{g} _{{\sss {P2}}}\right)^2L_{0}
+2(g_{{\sss {A}}}-\overline{g} _{{\sss {P2}}})
(\overline{g} _{{\sss {A}}}+\overline{g} _{{\sss {P1}}})L_{40}
+(\overline{g} _{{\sss {A}}}+\overline{g} _{{\sss {P1}}})^2L_{4}&;\mbox{for}~\mu=0 \\
  \left(g_{{\sss {A}}}+\mu \overline{g} _{{\sss {W}}}\right)^2 L_{\mu}&;\mbox{for}~\mu=\pm 1
\end{array}\right.,
\label{13}\er
are the effective coupling constants, which contain the lepton traces
\br
L_{4}&=&1+\frac{\pb\cdot\qb}{E_\ell E_\nu};~
L_{40}=\left(\frac{q_0}{E_\nu}+\frac{p_0}{E_\ell}\right),
\nn\\
L_{0}&=&1+\frac{2q_0p_0-\pb\cdot\qb}{E_\ell E_\nu};~
L_{\pm1}=1-\frac{q_0p_0}{E_\ell E_\nu}\pm
\left(\frac{q_0}{E_\nu}-\frac{p_0}{E_\ell}\right),
\label{14}\er
with
\br
q_0&=&{\kh}\cdot \qb=\frac{E_\nu(|\pb|\cos\theta-E_\nu)}{|\kb|};~
p_0={\kh}\cdot \pb=\frac{|\pb|(|\pb|-E_\nu\cos\theta)}{|\kb|}.
\label{15}\er
and   the momentum transfer $\kb$ is along the $z$ axis
($\kh\equiv \zh\equiv {\epsilon_0})$.

In going from the results for the neutrino-nucleus reaction cross section to that for
the muon capture rate one should keep in mind
that: i) the roles of $p$ and $q$ are interchanged within  the matrix element of
the leptonic current, which brings in a minus sign in the last term of $L_{\pm1}$,
  ii) the momentum transfer turns out to be $k=q-p$, and therefore the signs on the
  right hand sides of \rf{15} have to be  changed, and iii) the threshold values ($\pb\go 0: \kb\go
\qb, k_0\go E_\nu-m_\ell$) must be used for the lepton traces. All this yields:
\be
L_{4}=L_{40}=L_{0}=1;~~L_{\pm 1}= 1\mp 1.
\label{16}\ee
Finally, one should remember that instead of summing
over the initial lepton spins $s_\ell$, as done in \rf{10}, one
has now to average on the same quantum number.
 The resulting
transition amplitude $\T_{\Lambda}(J_f)$ is again of the form
\rf{11} but the effective charges are here:
\br
K_{{\sss{V}}}(\pb\go 0)&=&(g_{{\sss {V}}}+\overline{g} _{{\sss {V}}})^2
\nn\\
K_{{\sss {A}}}^{\mu}(\pb\go 0)&=&\del_{|\mu|, 1}
\left(g_{{\sss {A}}}-\overline{g} _{{\sss {W}}}\right)^2
+\del_{\mu, 0} \left(g_{{\sss {A}}}+\overline{g} _{{\sss
{A}}}-\overline{g} _{{\sss {P}}}\right)^2,
\label{17}\er
with
\br
\gv&=&\gV\frac{E_\nu}{2M};~\ga=\gA\frac{E_\nu}{2M};~
\gw=(\gV+\gM)\frac{E_\nu}{2M};~ \gp=\gpb-\gpa=\gP\frac{E_\nu}{2M}.
\label{18}\er
For the capture rate one gets \cite{Wal95}
\br
\Lambda(J_f)&=&\frac{E_\nu^2}{2\pi}|\phi_{1S}|^2\T_{\Lambda}(J_f),
\label{19}\er
where $\phi_{1S}$ is the  muonic bound state wave
function evaluated at the origin. Note that the neutrino energy is
fixed by the energy of the final state through the relation:
$E_\nu=m_\mu-(m_n-m_p)-E_B^\mu-E_f+E_i$, where $E_B^\mu$ is the
binding energy of the muon in the $1S$ orbit.

Lastly, we mention that the $B$-values for  the GT beta transitions are
defined and related to the ft-values as \cite{Tow95}:
\br
\frac{|g_{\sss A}\Bra{J_f}\s\Ket{J_i}|^{2}}{2J_i+1}\equiv
B(GT)&=&\frac{6146}{ft}~\mbox{sec}.
\label{20}\er

To start the discussion on the difficulties found by Volpe {\it et al.} \cite{Vol00},
it should be remembered that, the   pn-QRPA yields
the same energy spectra for the four  $(Z\pm 1,N\mp1)$ and $(Z\pm 1,N\pm 1)$ nuclei,
when  the BCS equations are solved in the parent
$(Z,N)$ nucleus under the constraint
\be
\sum_{k=n(p)}(2j_k+1)v_{j_k}^2=N (Z).
\label{21}\ee
This is a physically sound zero order approximation when the nuclei in question are far from
the closed shells and possess a significant neutron excess. Yet, as we show below,
the use of the QRPA in $N=Z$ nuclei is not free from care.

Let us define the quasiparticle energies relative to the Fermi levels:
\br
{E}_{j_k}^{(\pm)}&=&\pm E_{j_k}+ \l_{k};~~k=p,n,
 \label{22}\er
where $E_{j_k}$ and $\l_{k}$ are the BCS quasiparticle energies and
chemical potentials,
respectively.
In the particle-hole limit the energies ${E}_{j_k}^{(+)}$
(${E}_{j_k}^{(-)}$)
correspond to the  single-particle (-hole) excitations   for the levels
above (below) the Fermi surface \cite{Con84},  and to the 2p1h (1p2h)
seniority-one excitations  for levels below (above) the Fermi surface.
In nuclei with large neutron excess ${E}_{j_p}^{(\pm)}$ and
${E}_{j_n}^{(\pm)}$  are in general  quite different, but in $N=Z$ nuclei
the proton and neutron spectra are almost equal, except for the Coulomb energy displacement.
 As a consequence the unperturbed QRPA energies:
\br
\E_{j_pj_n'}&=& \left\{ \begin{array}{c}
  {E}_{j_p}^{(+)}-{E}_{j_n'}^{(-)};~\mbox{ for}:~(Z+1,N-1)  \\
- {E}_{j_p}^{(+)}+{E}_{j_n'}^{(-)};\mbox{ for}:~(Z-1,N+1) \\
 {E}_{j_p}^{(+)}+{E}_{j_n'}^{(-)};~~\mbox{ for}:~(Z+1,N+1)  \\
 - {E}_{j_p}^{(+)}-{E}_{j_n'}^{(-)};\mbox{ for}:~(Z-1,N-1)
\end{array}  \right.,
\label{23}\er
are almost degenerate with $\E_{j_p'j_n}$ \ie $\E_{j_pj_n'}\cong \E_{j_p'j_n}$,
for all four odd-odd $(Z\pm 1,N\mp1)$ and $(Z\pm 1,N\pm 1)$ nuclei.
Moreover, in the case of ${^{12}C}$, both  the proton and the neutron Fermi
energies are placed almost in the middle between the $1p_{3/2}$ and $1p_{1/2}$ shells.
This causes an additional degeneracy, namely $E_{1p_{3/2}}\cong E_{1p_{1/2}}$, resulting in
\be
\E_{{3/2},{1/2}}\cong\E_{{1/2},{3/2}}\cong\E_{{3/2},{3/2}}\cong\E_{{1/2},{1/2}},
\label{24}\ee
for ${^{12}N}$, ${^{12}B}$, ${^{14}N}$ and ${^{10}B}$.
But we know that the  physically sound energy sequences are:
\br
&{^{12}N}:&\E_{{1/2},{3/2}}(1p1h)<\E_{{3/2},{3/2}}(2p2h)
\simeq\E_{{1/2},{1/2}}(2p2h)<\E_{{3/2},{1/2}}(3p3h),
\nn\\
&{^{12}B}:&\E_{{3/2},{1/2}}(1p1h)<\E_{{3/2},{3/2}}(2p2h)
\simeq\E_{{1/2},{1/2}}(2p2h)<\E_{{1/2},{3/2}}(3p3h),
 \nn\\
&{^{14}N}:&~~\E_{{1/2},{1/2}}(2p)~<\E_{{1/2},{3/2}}(3p1h)
\simeq\E_{{3/2},{1/2}}(3p1h)<\E_{{3/2},{3/2}}(4p2h),
\nn\\
&{^{10}B}:&~~\E_{{3/2},{3/2}}(2h)~<\E_{{3/2},{1/2}}(1p3h)
\simeq\E_{{1/2},{3/2}}(1p3h)<\E_{{1/2},{1/2}}(2p4h),
\label{25}\er
as  can be easily seen from the scrutiny of the particle-hole limits
 for the seniority-two pn-states, which are indicated  parenthetically in \rf{25}.
The RPA correlations are unable to remedy the situation  and the
degeneracy \rf{24} among four lowest $\E_{pn}$ is the cause
for  the problems found in Ref. \cite{Vol00} regarding the ground state of $^{12}N$.

Well aware of all these difficulties,  Cha \cite{Cha83}, in his study of
the Gamow-Teller (GT) resonances, has solved the BCS equations in the daughter
nuclei under the constraints
\be
\sum_{k=n(p)}(2j_k+1)\tilde{v}_{j_k}^2=N\pm 1 ~(Z\mp 1),
\label{26}\ee
which gives way to the energy orderings \rf{25}.
Thus, the  problem risen by Volpe {\it et al} \cite {Vol00}  can, in principle, be solved by
using the Cha's recipe. However, the price to pay is that a different
QRPA equation has to be worked out for each nucleus separately, \ie one for the
$(Z+1,N-1)$ nucleus and the other for the $(Z-1,N+1)$ nucleus,
being in each case significant only the positive energy frequencies.
This means that we have to abandon the nice properties of the particle-hole
charge-exchange RPA, where: (1) only one RPA equation is solved for
the $(Z\pm 1,N\mp 1)$ nuclei, and
(2) both the positive and negative solutions are physically meaningful,
with  the $\beta^{+}$ spectrum viewed as an extension
of the $\beta^{-}$ spectrum to negative energies \cite{Boh75,Lan80,Krm80}.
Note also  that, in order to fulfill the GT sum rule, Cha has evaluated
the transition probabilities with the usual pairing factors
$u$'s and $v$'s, obtained from \rf{21}.
None of the  undesirable features  of the Cha's method  appear within
the charge-exchange PQRPA. This approach has been presented in detail in
Ref.  \cite{Krm93}, and we just mention here that  the PBCS quasiparticle energies read:
\br
{E}_{j}^{(+)}&=&\frac{R_0^K(j)+R_{11}^K(jj)}{I^K(j)}-\frac{R_0^{K}}{I^K},
\nn\\
{E}_{j}^{(-)}&=&-\frac{R_0^{K-2}(j)+R_{11}^{K-2}(jj)}{I^{K-2}(j)}+\frac{R_0^{K-2}}{I^{K-2}},
\label{27}\er
where $K=N,Z$ and the quantities $R^K$ and $I^K$ are defined in \cite{Krm93}.

\begin{table}[h]
\caption{ BCS and PBCS results for neutrons. $E^{exp}_j$ stand for the experimental
energies used in the fitting procedure, and $e_j$ are the resulting s.p.e.
The underlined   quasiparticle energies
correspond to single-hole excitations (for $j=1s_{1/2},1p_{3/2}$)  and to
single-particle excitations (for $j=1p_{1/2},
1d_{5/2},2s_{1/2},1d_{3/2},1f_{7/2},2p_{3/2}$).
The non-underlined energies are   mostly two hole-one particle and two
particle-one hole excitations. The fitted values of the pairing strengths
$v^{pair}_s$ in units of MeV-fm$^3$ are also displayed.}
\begin{center}
\label{table:1}
\newcommand{\cc}[1]{\multicolumn{1}{c}{#1}}
\renewcommand{\tabcolsep}{0.9pc} 
\renewcommand{\arraystretch}{1.2} 
\bigskip
\begin{tabular}{|c| r| rrr|rrr|}\hline
$$&$$&$$&$BCS$&$$&$$&$PBCS$&$$
\\
$Shell$&$E^{exp}_j$
&$E_j^{(+)}$&$E_j^{(-)}$&$e_j$&$E_j^{(+)}$&$E_j^{(-)}$&$e_j$
\\ \hline
$1s_{1/2}$&&$11.34$&$\underline{-35.13}$&$-23.58$  &$
19.93$&$\underline{-34.99}$& $-22.37$
\\
$1p_{3/2}$&$-18.72$&$-5.07$&$\underline{-18.72}$&$-7.80$ &
$-1.28$&$\underline{-18.73}$& $-7.24$
\\
$1p_{1/2}$&$-4.94$&$\underline{-4.94}$&$-18.85$&$-2.07$
&$\underline{-4.95}$&$-22.33$&$-1.51$
\\
$1d_{5/2}$&$-1.09$&$\underline{-1.09}$&$-22.70$&$2.12$
&$\underline{-1.09}$&$-26.82$ & $2.16$
\\
$2s_{1/2}$&$-1.85$&$\underline{-1.86}$&$-21.93$&$2.70$
&$\underline{-1.85}$&$-25.98$ & $2.68$
\\
$1d_{3/2}$&$2.72$&$\underline{2.72}$&$-26.51$ &$6.24$
&$\underline{2.73}$&$-30.79$  & $6.26$
\\
$1f_{7/2}$&$5.81$&$\underline{5.82}$&$-29.61$ &$8.14$
&$\underline{5.83}$&$-33.61$  & $8.17$
\\
$2p_{3/2}$&$7.17$&$\underline{7.18}$&$-30.98$ &$11.49$
&$\underline{7.16}$&$-35.23$  & $11.47$
\\
$2p_{1/2}$&&$\underline{12.89}$&$-36.69$&$17.30$
&$\underline{12.89}$&$-41.01$ & $17.32$
\\
$1f_{5/2}$&&$\underline{16.72}$&$-40.52$&$19.18$
&$\underline{16.72}$&$-44.58$ & $19.21$
\\\hline
$v_{{s}}^{{pair}}$&&& &$23.16$& &&$23.92$
\\\hline
\end{tabular}
\end{center}
\end{table}

The numerical calculations were performed   within the $nl=(1s,1p,1d,2s,1f,2p)$
configuration space, and  for the residual interaction we adopted  the delta force
\br
 V &=&-4 \pi \left(v_sP_s+v_tP_t\right)\delta(r),
 \label{28} \er
where $v_s$ and $v_t$ are given in units of MeV-fm$^3$.

The  bare single-particle energies (s.p.e.) $e_j$ were fixed from the experimental
energies of the odd-mass nuclei  $^{11}C$, $^{11}B$, $^{13}C$ and $^{13}N$.
That is, we  assume that the ground states in $^{11}C$ and $^{11}B$
are pure quasi-hole excitations ${E}_{1p_{3/2}}^{(-)}$, and that
the lowest observed ${1/2}^{-}, {5/2}^{+},{1/2}^{+},{3/2}^{+}, \\{7/2}^{-}$ and
${3/2}^{-}$ states  in $^{13}C$ and $^{13}N$ are pure quasi-particle excitations
${E}_{j}^{(+)}$  with $j=(1p_{1/2},1d_{5/2},2s_{1/2},1d_{3/2},1f_{7/2}$,
$2p_{3/2})$. This is in essence the idea of the inverse-gap-equation (IGE) method \cite{Alz69},
which also fixes the value of  the singlet strength
within the pairing  channel ($v^{pair}_s$).
We have  considered the faraway orbitals $1s_{1/2}$, $2p_{1/2}$ and
$1f_{5/2}$ as well. Their s.p.e. were taken to by that of
a harmonic oscillator (HO) with standard parametrization.
The single-particle wave
functions were also approximated with that of the HO with the length parameter $b=1.67$
fm, which corresponds to the estimate $ \hbar \omega =45A^{-1/3}-25A^{-2/3}$~ MeV
for the oscillator energy.

The BCS and PBCS results for neutrons are displayed in Table 1.
The underlined  quasiparticle energies
correspond to single-hole  and  single-particle excitations,
while the non-underlined ones are basically 2h1p and 2p1h excitations.
Note that, while  the first ones  are
fairly similar within the BCS and PBCS approaches,  the last ones are quite
different. (The resulting  s.p.e. are also quite similar.)
Analogous results are obtained for protons, with the same value of $v_{{s}}^{{pair}}$.

The unperturbed energies $\E_{j_pj_n'}$  of  lowest four pn
quasiparticle states  within the BCS and PBCS approximations are shown in
Table 2. For comparison, the results obtained with the Cha's method are also
displayed in the same table. It is easy to see that, while the standard BCS
approximation exhibits the degeneracy \rf{24}, the Cha's recipe and the PBCS
approach produce the energy sequences \rf{25}, being the energy separations
between the 1p1h, 2p2h and 3p3h-like states
significantly larger in the later case. This does not take us by surprise
since the role of the number projection  is precisely that of restoring  the correct number
of particles and holes.

\begin{table}[h]
\caption{Unperturbed energies $\E_{j_pj_n'}$ (in units of MeV) of  lowest
four proton-neutron quasiparticle states in the neighborhood of ${^{12}C}$, within the
approximations:
(a) BCS equations are solved in ${^{12}C}$ with the condition  \rf{21},
(b) BCS equations are solved in daughter nuclei, employing \rf{24} as
suggested by  Cha \protect\cite{Cha83}, and (c) number projected BCS (PBCS) equations
are employed.
The underlined energies are equal for all  three cases, because they are adjusted
to the experimental data via the IGP procedure \protect\cite{Alz69}.}
\label{table:2}
\newcommand{\cc}[1]{\multicolumn{1}{c}{#1}}
\renewcommand{\tabcolsep}{0.3pc} 
\renewcommand{\arraystretch}{1.2} 
\bigskip
\begin{tabular}{|c|c|ccc|ccc|ccc|ccc|}\hline
$ j_{p}j'_{n}$&$E_{{j_p}}+E_{{j'_n}}$&&
${^{12}N}$&&&${^{12}B}$&&&${^{14}N}$&&&${^{10}B}$&
\\
&&(a)              &(b)&(c)      &(a)&(b)&(c)&(a)&(b)&(c)&(a)&(b)&(c)\\
\hline
${1/2},{3/2}$&14.0&\underline{16.8}&\underline{16.8}&\underline{16.8}
&11.2&15.2&18.7&-7.0&-4.8&-3.2&35.0&36.7&38.5\\
${3/2},{1/2}$&13.8&16.6&20.6&23.8&\underline{11.0}&\underline{11.0}&\underline{11.0}&-7.2&-5.0&-3.5&34.8&36.6&38.3\\
${1/2},{1/2}$&14.1&16.9&18.7&20.4&11.3&13.1&14.8&
\underline{-6.9}&\underline{-6.9}&\underline{-6.9}&35.1&38.7&42.1\\
${3/2},{3/2}$&13.7&16.5&18.7&20.2&10.9&13.1&14.7&-7.3&-2.9&
0.2&\underline{34.7}&\underline{34.7}&\underline{34.7}
\\\hline
\end{tabular}
\end{table}

After having established truthful unperturbed PQRPA energies we proceed  with
 full calculations, where  the values of $v_s$ and $v_t$ within the
particle-particle ($pp$) and particle-hole ($ph$) channels are needed. In  similar calculations
of double beta decaying nuclei \cite{Bar98,Hir90}, which possess significant neutron excess,
the following procedure has been pursued:
(i) $v^{ph}_s$ and $v^{ph}_t$ were taken from the study of energetics  of
the GT resonances done  by Nakayama {\it et al.}  \cite{Nak82} (see also ref. \cite{Cas87}),
and (ii) the $pp$ strengths  were fixed on the basis of the isospin and SU(4) symmetries as:
$v^{pp}_s\equiv v^{pair}_s$, and $v^{pp}_t\gsim v^{pp}_s$.
Different to what happens in the $N> Z$ nuclei, such a parametrization is
not suitable for the $N=Z$ nuclei, and the best agreement with data is obtained when
the $pp$ channel is totally switched  off. Thus we will exhibit here only the results for
$v^{pp}_s=v^{pp}_t=0$, and the next three sets of  $ph$ parameters:

{\it Calculation I}:  $ v^{ph}_s=v^{pair}_s=23.92$ MeV-fm$^3$, and
 $ v^{ph}_t=v^{ph}_s/0.6=39.86$ MeV-fm$^3$. That is, the singlet ph strength is taken
 to be the same as $v^{pair}_s$ obtained from the gap equation,
while the triplet $ph$ depth is estimated from the relation
used by Goswami and Pal \cite{Gos62} in the RPA calculation of ${^{12}C}$.

{\it Calculation II}: $v^{ph}_s=27$  MeV-fm$^3$, and $v^{ph}_t=64$ MeV-fm$^3$. These values
were suggested in refs.  \cite{Nak82,Cas87} and have been  extensively used in the QRPA
calculations of ${^{48}Ca}$ \cite{Bar98,Hir90}.

{\it Calculation III}: $v^{ph}_s=v^{ph}_t=45$ MeV-fm$^3$. This parametrization gives
fairly good results for the energies of the $J^\pi=0_1^+$ and  $1_1^+$ states in
${^{12}B}$ and ${^{12}N}$.
\begin{table}[h]
\caption{Results for the energy of the  $J^{\pi}=1^{+}_1$ state in
${^{12}N}$ in units of MeV, the average $B(GT)$-value for the $\beta$-decay from  ${^{12}N}$
and ${^{12}B}$, and  the $\mu$-capture rates  to the  ground state ($\Lambda^{exc}$)
and to all final states ($\Lambda^{inc}$) in ${^{12}B}$
in units of $10^3$~sec$^{-1}$. In the upper part of the table the smallest (largest)
estimates obtained in  previous RPA calculations  are  shown.
As explained in the text three different PQRPA
calculations are presented. The lower and upper experimental $B(GT)$-value correspond to
${^{12}N}$ and ${^{12}B}$, respectively. }
\label{table:3}
\newcommand{\cc}[1]{\multicolumn{1}{c}{#1}}
\renewcommand{\tabcolsep}{0.9pc} 
\renewcommand{\arraystretch}{1.2} 
\bigskip
\begin{tabular}{|l|lccc|}\hline
&$E({1^{+}_1})$&$B(GT)$&$\Lambda^{exc}$&$\Lambda^{inc}$\\
 \hline RPA$^{\cite{Kol94}}$&&$1.94~(2.02)$&$22.8~(25.4)$&$57~(59)$\\ RPA$^{
\cite{Aue97}}$&&$$&$32.4~(34.8)$&$69~(72)$\\ RPA+pair$^{
\cite{Aue97}}$&&&$4.1~(7.3)$&$31~(36)$\\
CRPA$^{\cite{Kol99}}$&&0.693~(0.776)&$8.5~(9.3)$&$40~(42)$\\
RPA$^{\cite{Vol00}}$&$13.74$&$2.03$&$25.4$&$51$\\\hline
PBCS&$16.78$&$1.063$&$15.2$&$66$\\
PQRPA~~~(I)&$17.89$&$0.568$&$7.8$&$46$\\
PQRPA~~(II)&$18.14$&$0.477$&$6.5$&$40$\\
PQRPA~(III)&$18.13$&$0.480$&$6.5$&$42$\\
\hline
Expt.&$17.34^{\cite{Ajz85}} $&$0.466-0.526^{\cite{Al78}}$&$6.2 \pm
0.3^{\cite{Mill72}}$&$38 \pm 1^{\cite{Suz87}}$
\\\hline
\end{tabular}
\end{table}

In Table 3 we  confront  our PBCS and PQRPA results with previous RPA
calculations \cite{Kol94,Aue97,Kol99,Vol00}, and with experiments
\cite{Ajz85,Al78,Mill72,Suz87} for: the energy of the $J^{\pi}=1^{+}_1$ state in ${^{12}N}$,
the $B(GT)$-value for the $\beta$-decay from  ${^{12}N}$ and ${^{12}B}$, and  the
exclusive and inclusive $\mu$-capture rates  to ${^{12}B}$:
$\Lambda(J^{\pi}_f=1^{+}_1)$ and $\sum_{J^{\pi}_f}\Lambda(J^{\pi}_f)$.
 We don't show our QRPA results because of the above mentioned difficulties  with the
 $J^{\pi}=1^{+}_1$ ground states in ${^{12}N}$ and ${^{12}B}$.
In comparing the calculations of $B(GT)$ with data it should be remembered that it
is still not  clear the origin of the observed $10\%$ difference measured  $ft$ values
for the GT $\beta$-decays from the ground states $J^\pi=1^+$ in ${^{12}B}$ and ${^{12}N}$
to the ground state $J^\pi=0^+$  in ${^{12}C}$:
$ft({^{12}B})= (1.1669\pm0.0037) \x 10^{4}\mbox{seg}$, and
$ft({^{12}N})= (1.3178\pm0.0084) \x 10^{4}\mbox{seg}$ \cite{Al78}.
In the past this difference has been  attributed mostly to the violation of charge
symmetry in the involved nuclear states, and occasionally also to the second class
current (or induced tensor interaction) which violates the G-parity \cite{Bli65,Ema72,Hol76}.
\footnote{Presently, the study of the G-parity irregular weak nucleon
current is still of interest \cite{Shi96,Min01}.}
 As this kind of effects are not considered  in the present work the above $ft$ values
 will be taken as lower and upper experimental limits.  The corresponding
$B$-values, obtained  from  \rf{20} ($B_B(GT)=0.526$ and $B_N(GT)=0.466$)  are shown
in Table 3. Due to the same reason, the small difference ($\lsim ~3\%$) between
the theoretical results for $B_B(GT)$ and $B_N(GT)$ is not physically relevant and
only the mean values $(B_B(GT)+B_N(GT))/2$ are exhibited.

\begin{table}[h]
\caption{Results for  averaged exclusive
and  inclusive neutrino-nucleus cross sections $\langle \s_{e} \rangle$
(in units of $10^{-42} $~cm$^{2}$) and $\langle \s_{\mu} \rangle$
(in units of $10^{-40} $~cm$^{2}$). (See caption to Table 1.)}
\label{table:4}
\newcommand{\cc}[1]{\multicolumn{1}{c}{#1}}
\renewcommand{\tabcolsep}{0.1pc} 
\renewcommand{\arraystretch}{1.2} 
\bigskip
\begin{tabular}{|l|cccc|}\hline
&$\langle \s_{e}^{exc} \rangle$&$\langle \s_{e}^{inc} \rangle$
&$\langle \s_{\mu}^{exc} \rangle$&$\langle \s_{\mu}^{inc} \rangle$\\
\hline
 RPA$^{\cite{Kol94}}$&$36.0~(38.4)$&$42.3~(44.3)$&$2.48~(3.11)$&$21.1~(22.8)$\\
 RPA$^{\cite{Aue97}}$&$54.8~(68.2)$&$63.2~(76.3)$&$3.35~(3.80)$&$21.1~(22.4)$\\
 RPA+pair$^{\cite{Aue97}}$&$7.1~(16.0)$&$12.9~(22.7)$&$0.39~(0.77)$&$13.5~(15.2)$\\
CRPA$^{\cite{Kol99}}$&$12.5~(13.9)$&$18.15~(19.28)$&$1.06~(1.06)$&$17.8~(18.2)$\\
RPA$^{\cite{Vol00}}$&$50.0$&$55.1$&$2.09$&$19.2$\\
QRPA$^{\cite{Vol00}}$&$42.9$&$52.0$&$1.97$&$20.3$\\\hline
PBCS&$21.0$&$41.2$&$1.67$&$19.1$\\
PQRPA~~~(I) &$9.9$&$21.6$&$0.72$&$14.6$\\
PQRPA~~(II) &$8.0$&$18.5$&$0.56$&$12.8$\\
PQRPA~(III) &$8.1$&$17.4$&$0.56$&$13.4$\\
\hline
Expt.&$9.1 \pm 0.4\pm 0.9 ^{\cite{LSND97}}$&$14.8 \pm0.7\pm 1.4^{\cite{LSND97}}$
&$0.66\pm 0.1\pm0.1^{\cite{LSND97b}}$&$12.4 \pm 0.3 \pm 1.8^{\cite{LSND97b}} $\\
&$8.9 \pm 0.3\pm 0.9 ^{\cite{LSND01}}$&$13.2 \pm0.4\pm 0.6^{\cite{LSND01}}$
&$0.56\pm 0.08\pm0.10^{\cite{LSND02}}$&$10.6\pm 0.3 \pm 1.8^{\cite{LSND02}} $\\
\hline
\end{tabular}
\end{table}

Similarly,  in Table 4 are given the  results for  the exclusive and inclusive
flux-averaged  neutrino
scattering cross sections to ${^{12}N}$: $\langle \s_\ell(J^{\pi}_f=1^{+}_1) \rangle$,
$\sum_{J^{\pi}_f}\langle \s_\ell(J^{\pi}_f) \rangle$ with $\ell=e,\mu$.
They are defined as
\be
\langle \s_\ell(J_{f}) \rangle= \int dE_{\nu}
\s(E_\ell=E_i-E_f-E_\nu,J_{f}) \bar{f}(E_{\nu}),
\label{29}\ee
where $\bar{f}(E_{\nu})$ is the normalized  neutrino flux. For electron neutrinos
it was approximated by the Michel spectrum, and for the muon neutrinos we used that from
ref. \cite{LSND}.

\begin {table}[h]
\begin{center}
\caption {Results for the energies (in units of MeV) and the partial muon capture rates
(in units of $10^3$~s$^{-1}$) the bound excited states in $^{12}B$.  In the upper part of
the table are  shown the previous theoretical calculations based on the
 RPA    \protect\cite{Kol94,Kol94a} (where only the results for $\Lambda$ are reported)
  and the on  shell model  \protect \cite{Aue02}.} \label{table:5}
\newcommand{\cc}[1]{\multicolumn{1}{c}{#1}}
\renewcommand{\tabcolsep}{0.3pc} 
\renewcommand{\arraystretch}{1.2} 
\bigskip
\begin{tabular}{| c c|c|c|c|c|}\hline
Model            &$J^{\pi}_f$&$1^+_1$&$2^+_1$&$2^-_1$&$1^-_1$\\\hline\hline
RPA$^{\cite{Kol94,Kol94a}}$  &$\Lambda$&$25.4~(22.8)$&$\leq
10^{-3}$&$0.04~(0.02)$&$0.22~(0.74)$\\
SM$^{\cite{Aue02}}$  &E         & $0.00$& $0.76$& $1.49$& $1.99$\\
                 &$\Lambda$ & $6.0 $& $0.25$& $0.22$& $1.86$\\\hline
PBCS             &E         & $0.00$& $0.00$& $3.10$& $3.10$\\
                 &$\Lambda$ & $15.4$& $0.40$& $1.70$& $1.13$\\
PQRPA~~~(I)      &E         & $0.00$& $0.34$& $2.83$& $3.13$\\
                 &$\Lambda$ & $7.83$& $0.21$& $0.34$& $0.66$\\
PQRPA~~(II)      &E         & $0.00$& $0.50$& $2.82$& $3.31$\\
                 &$\Lambda$ & $6.50$& $0.16$& $0.18$& $0.51$\\
PQRPA~(III)      &E         & $0.00$& $0.28$& $2.82$& $2.97$\\
                 &$\Lambda$ & $6.54$& $0.17$& $0.18$& $0.58$\\\hline
Expt$^{\cite{Mae01,Sto02}}$&E         & $0.00$& $0.95$& $1.67$& $2.62$\\
                 &$\Lambda$ & $6.00\pm 0.40$& $0.21\pm 0.10$& $0.18\pm
0.10$& $0.62\pm 0.20$\\
\hline\hline
\end{tabular}
\end{center}
\end {table}
We wish to restate the ingredients that play a part in the agreement
between the data and calculations for the ground state  processes within the triad
$\{{{^{12}B},{^{12}C},{^{12}N}}\}$. They are: (a) the pairing short range correlations,
which are added to improve the description of the $^{12}C$ ground state,
(b) the RPA-type correlations, which are repulsive in the particle-hole channel,
and (c) the effective axial-vector
coupling constant $\gA=1$, which in principle simulates the removal of the spin
strength due to the coupling to the $\Delta$ resonance
\cite{Bro85,Cas87,Ost92,Aue02}. For instance, these effects reduce the bare single-particle
value  $B(GT)=(16/9)\gA^{2}$
by factors $1.7$, $1.8-2.2$ and $1.6$, respectively.
It is worthy of note that the PBCS by itself reproduces better the
data  than the majority of  previous
RPA and QRPA calculations \cite{Kol94,Aue97,Kol99,Vol00}. We have considered
 all orbitals from
$1s_{1/2}$ up to $1f_{5/2}$, but the valence
$p$-shell correlations (both  pairing and  RPA like) are
definitely the most important ones for the quenching of the $1_1^{+}\leftrightarrow 0_1^{+}$
transition rates. Yet, as discussed by Vogel \etal \cite{Kol99,Vog99}, the effect
of these correlations on the dipole and quadrupole operators is very tiny.

In addition to the total $\mu$ capture rates in Table 3, we show the individual rates
to the individual bound states of ${^{12}B}$ in Table 5. They represent another test for
our calculation and have been derived from the intensities of the observed
de-excitation  $\gamma$ rays following the $\mu$ capture \cite{Mae01,Sto02}.
The agreement between the experiment and the present PQRPA estimate for
the energies of the $J^\pi_f=2^+_1,2^-_1$ and $1^-_1$ states is only moderate, but that
for the capture rates is as good or even better than in a recent shell model (SM) study
\cite{Aue02}.

In summary, we have shown that  to account for the weak decay
observables  around ${^{12}C}$ in the framework of the RPA, besides including
 the BCS correlations, it is imperative to perform the particle
number projection, and this is the way out of the RPA puzzle in ${^{12}C}$.
More, as far as we are acquainted with, such an important effect
of the projected linear response theory for charge-exchange excitations
has never before been observed, indicating that it
could be more relevant in  light $N=Z$ nuclei than in heavy nuclei with large neutron
excess \cite{Krm93}.
Thus, it could be interesting enough to investigate the consequences of the PQRPA in other
light $N=Z$ and   $N\cong Z$ nuclei.
 On the other hand, the fact that we have been forced to
 switch off completely the residual interaction in the particle-particle channel could
 indicate that some relevant piece of physics is still lacking in our approach.
 In this sense it would be very illuminating to redo the PQRPA calculations
with more realistic forces than the one used here.

\bigskip
The authors acknowledge  the support of   ANPCyT (Argentina) under
grant BID 1201/OC-AR (PICT 03-04296) and of CONICET under grant PIP 463.
 F.K. and A.M. are fellows of the CONICET  Argentina.

\bigskip


\begin{thebibliography}{9}
\bibitem{Kol94} E. Kolbe, K. Langanke and S. Krewald,  Phys. Rev. {\bf C 49}, 1122 (1994).
\bibitem{Aue97} N. Auerbach, N. Van Giai, O.K. Vorov,  Phys. Rev. {\bf C 56}, 2368 (1997).
\bibitem{Kol99} E. Kolbe, K. Langanke and P. Vogel, Nucl. Phys.  {\bf A 652}, 91 (1999).
\bibitem{Vol00} C. Volpe, N. Auerbach, G. Col\`o, T. Suzuki, N. Van Giai, Phys. Rev. {\bf C 62},
015501 (2000).
\bibitem{Krm93} F. Krmpoti\'c, A. Mariano, T.T.S. Kuo, and K. Nakayama,  Phys. Lett.
{\bf B 319} 393(1993).
\bibitem{Bar98} C. Barbero, F. Krmpoti\'c, and D. Tadi\'c,
Nucl. Phys. {\bf A 628}, 170 (1998);
 C. Barbero, F. Krmpoti\'c, A. Mariano and D. Tadi\'c, Nucl. Phys. {\bf A 650}, 485 (1999).
\bibitem{Tom91} T. Tomoda, {\it Rep. Prog. Phys.} {\bf {54}} (1991) 53.
\bibitem{Doi93} M. Doi and T. Kotani, {\it Prog. Theor. Phys.} {\bf {89}}, (1993) 139.
\bibitem{Tow95}I.S. Towner and  J.C. Hardy, {\it The Nucleus as a Laboratory for
Studying Symmetries and Fundamental Interactions}, eds. E.M. Henley and
W.C. Haxton, nucl-th/9504015.
\bibitem{Bro85} B.A. Brown and B.H. Wildenhal, At. Data Nucl. Data Tables
{\bf 33}, 347 (1985).
\bibitem{Cas87} H. Castillo and F. Krmpoti\'c, Nucl. Phys. {\bf A 469}, 637
(1987), and references there in.
\bibitem{Ost92} F. Osterfeld, Rev. Mod. Phys. {\bf 64}, 491 (1992).
\bibitem{Kur90} T. Kuramoto, M. Fukigita, Y. Kohyama and K. Kubodera,
{\it Nucl. Phys.} {\bf A 512} (1990) 711.
\bibitem{Luy63} J.R.Luyten, H.P.C. Rood and H.A. Tolhoek,
  Nucl.Phys. {\bf 41},236 (1963).
\bibitem{Kuz01}  V.A. Kuzmin, T.V. Tetereva, K. Junker, A.A. Ovchinnikova,
nucl-th/0104061.
\bibitem{Bli66} R.J. Blin-Stoyle and S.C.K. Nair, Adv. Phys. {\bf 15},493 (1966).
\bibitem{Wal95} J.D. Walecka, {\it Theoretical Nuclear and Subnuclear Physics},
{\it Oxford University Press, New York}, 531 (1995).
\bibitem{Con84} C. Conci, V. Klempt and J. Speth,
Phys. Lett. {\bf 148B} (1984) 405; C. Conci, Ph.D. Thesis, unpublished, J\"ulich 1984.
\bibitem{Cha83} D. Cha, Phys. Rev. {\bf C 27}, 2269 (1983).
\bibitem{Boh75} A. Bohr B. R. Mottelson, {\it Nuclear Structure} {\bf
Vol.II} (W.A. Benjamin Inc., New York, Amsterdam, 1975).
\bibitem{Lan80} A.M. Lane and J. Martorrel, Ann. Phys. {\bf 129},273
(1980).
\bibitem{Krm80} F. Krmpoti\'c, K. Ebert, W. Wild, Nucl. Phys. {\bf A 342},
497 (1980).
\bibitem{Alz69} R. Alzetta, T. Weber, Y. K. Gambhir, M. Gmitro, J.
Sawicki, and A. Rimini     Phys. Rev. {\bf 182} , 1308(1969)
\bibitem{Hir90} J. Hirsch and F. Krmpoti\'c, Phys. {\bf Rev.C 41}, 792(1990);
F. Krmpoti\'c and Shelly Sharma, Nucl. Phys. {\bf A 572}, (1994) 329.
\bibitem{Nak82} K. Nakayama, A. Pio Gale\~{a}o and F. Krmpoti\'{c},
Phys. Lett. {\bf B 114} (1982) 217.

\bibitem{Gos62} A. Goswami and M.K. Pal, Nucl. Phys. {\bf 35}, (1962) 544.
\bibitem{Ajz85} F. Ajzenberg-Selove, Nucl. Phys. {\bf A 433}, 1(1985) .
\bibitem{Al78} D. E. Alburger and A.M. Nathan, {\it Phys. Rev.} {\bf C 17} (1978) 280.
\bibitem{Mill72} G. H. Miller {\it et al., Phys.  Lett.} {\bf B 41}, (1972) 50.
\bibitem{Suz87} T. Suzuki {\it et al., Phys.  Rev.} {\bf C 35}, (1987) 2212.
\bibitem{Bli65} R.J. Blin-Stoyle and M. Rosina, Nucl. Phys. {\bf 70}, 321
(1965).
\bibitem{Ema72} B. Eman, D. Tadi\'c, F. Krmpoti\'c and L. Szybisz, Phys. Rev. {\bf C 6}, 1
(1972).
\bibitem{Hol76} B.R. Holstein and S.B. Trieman, Phys. Rev.{\bf C 13}, 3059
(1978).
\bibitem{Shi96} H. Shiorui,  Nucl. Phys. {\bf A 603}, 281 (1996).
\bibitem{Min01} K. Minamisono, {\it et al} Phys. Rev. {\bf C 65}, 015501
(2001).
\bibitem{LSND97} LSND Collaboration, C. Athanassopulus et al., Phys. Rev.{\bf C 55}, 2078 (1997).
\bibitem{LSND97b} LSND Collaboration, C. Athanassopulus et al., Phys. Rev.{\bf C 56}, 2806
(1997).
\bibitem{LSND01} LSND Collaboratio, L-B. Auerbach et al., Phys. Rev.{\bf C 64}, 065501 (2001).
\bibitem{LSND02} LSND Collaboratio, L-B. Auerbach et al., arXiv:nucl-ex/0203011.
\bibitem{LSND} LSND home page, http://www.neutrino.lanl.gov/LSND/figures.
\bibitem{Kol94a} E. Kolbe, K. Langanke and P. Vogel,  Phys. Rev. {\bf C 50}, 2576 (1994).
\bibitem{Aue02} N. Auerbach and B.A. Brown, Phys. Rev. {\bf C 65}, 024322 (2002).
\bibitem{Mae01} D.F. Maesday, Phys. Rep. {\bf 354}, 243 (2001).
\bibitem{Sto02}T.J. Stocki, D.F. Maesday, E. Gete, M.A. Saliba, and T.P. Gorrinde Nucl. Phys.
{\bf A 697}, 55 (2002).
\bibitem{Vog99} P. Vogel, nucl-th/9901027.

\end{thebibliography}
\end{document}